\documentclass[]{interact}

\usepackage[natbibapa,nodoi]{apacite}

\usepackage{color}

\usepackage{comment}
\usepackage{multirow}
\usepackage{float}
\usepackage{url}

\begin{document}

\articletype{}

\title{Teaching Business Process Modeling to Leverage Soft Skills of Computing Students}

\author{
\name{Maria Istela Cagnin Author\textsuperscript{a}\thanks{CONTACT Maria Istela Cagnin Author. Email:  istela.machado@ufms.br} and Rick Kazman\textsuperscript{b} and Elisa Yumi Nakagawa\textsuperscript{c}}
\affil{\textsuperscript{a} College of Computing (Facom), Federal University of Mato Grosso do Sul, Cidade Universidade, Campo Grande-MS, Brazil; \textsuperscript{b} The Shidler College of Business, University of Hawaii at Manoa, 2500 Campus Rd, Honolulu-Hawaii, United States \textsuperscript{c} Department of Computer Systems, University of São Paulo (USP), Trabalhador São-Carlense Avenue, 400, São Carlos-SP, Brazil}
}

\maketitle

\begin{abstract}
\textbf{Background and Context:} 
Organizations have highly depended on software systems to realize their business processes and achieve strategic business goals and competitive advantage; hence, computing professionals who can understand these processes could more effectively build such systems. At the same time, while Artificial Intelligence (AI) has increased the productivity in developing these and other systems, professionals should be better prepared due to the competitiveness of the job market; thus, soft skills could be an important factor to ensure their position in this market.
\textbf{Objective:}
This paper aims to analyze and present the impact of teaching business process modeling to leverage  key soft skills of future computing professionals.
\textbf{Method and Findings:} 
Our method involved an industry-driven, project-based approach with 53 computing students who actively interacted with stakeholders and modeled the organization's business process. We systematically collected the students' feedback, including the soft skills practiced and demonstrating that professionalism, initiative, motivation, communication, and teamwork can be leveraged.
\textbf{Implications:}
The main implication of this work for the Computer Science Education research community is to provide evidence that teaching business process modeling subjects is particularly beneficial for future professionals, as it could contribute to preparing them regarding their soft skills for the era of AI-assisted software development.
\end{abstract}


\section{Introduction}
\label{sec:introduction}
Software systems are increasingly crucial in organizations and  support the execution of business activities, including the decision-making process. These systems are often highly complex and involve diverse stakeholders from different organizational sectors. They are critical to help organizations achieve their business goals and obtain competitive advantage. Computing professionals must be prepared to develop these systems and for that it is essential they understand business level information: the business processes that are a valuable source for gathering software requirements \citep{kraupvsa2024}.

The ACM Education Board and the IEEE Computer Society Educational Activities Board have worked to mitigate the gap between the business and technical levels in undergraduate courses for software engineering \citep{ACM_SE_2014} and information systems~\citep{ACM_IS_2020}. To prepare future practitioners and meet the ACM/IEEE curricula recommendations, some universities have included  business process modeling in their undergraduate computing courses, employing not only theoretical content but also exercises \citep{recker2009teaching}, well-known specific cases \citep{bandara2020}, gamification \citep{garaccione2024}, token-based animations~\citep{maslov2024}, and flipped learning \citep{grzesiak2024}. Each of these pedagogical approaches contributes to improving the teaching and learning process. 

In parallel, it is notable an increasing adoption of Artificial Intelligence (AI) to support varied tasks in the software development (e.g, requirements structuring, architectural design, code generation, autonomous testing, predictive maintenance, and so forth) \citep{romero2023,durrani2024}. As a consequence, improvement in productivity has been observed~\citep{nguyen2024}, while the job market has become more competitive for new computing professionals~\citep{teng2024}.
In this context, good soft skills could become even more vital for improving employability~\citep{polakova2023,wilkinson2024,tes2025}. At the same time, teaching business process modeling also poses significant challenges for educators, as it requires hard skills from students, like using specific elements of business process modeling, modeling business processes, and mastering supporting tools. It also requires soft skills from students, such as teamwork and communication. We raise then the following research question: 
\textit{Which are the hard and soft skills undergraduate computing students experience when taught business process modeling, contributing to their training as future professionals}?

The goal of this paper is to analyze and present the impact of teaching business process modeling on key soft skills for future computing professionals. To this end, we adopted 
a method that involved an industry-driven, project-based learning approach to teach business process modeling to 53 undergraduate computing students at a large university. Students were immersed in organizations' environments and were required to increasingly understand their business processes and model them using BPMN\footnote{\url{https://www.bpmn.org/}}. From systematically collected students' feedback, we found that they were able to develop important soft skills: professionalism, initiative, motivation, communication, and teamwork. They also practiced hard skills---primarily how to use BPMN elements and supporting tools. The contribution of this paper is to make educators aware that 
an industry-driven, project-based approach, which can be easily incorporated into business process modeling subjects, offers students an opportunity to develop soft skills beyond what would be usually practiced in class-oriented subjects. We believe this paper could be helpful for those interested in exercising the soft skills of computing students through teaching business process modeling. 

The remainder of this paper is organized as follows. Section~\ref{sec:back} briefly presents the related work, regarding the  learning approaches adopted to teaching business process modeling and the essential soft skills in the era of AI-assistance software development. Section \ref{sec:report} 
presents our method and participants' profile. Section \ref{sec:results} discusses the results. Section \ref{sec:discussions} presents the findings, future works, and threats to validity. Finally, Section~\ref{sec:conclusion} concludes this work.

\section{Related Work}
\label{sec:back}
\subsection{Teaching Business Process Modeling}\label{sec:relatedwork}

Scientific studies on teaching business process modeling have explored diverse learning approaches, often blending them, to obtain better results in the teaching-learning process. Among recent learning approaches, studies have focused on cases \citep{bandara2020}, gamification \citep{garaccione2024}, token-based animations~\citep{maslov2024}, flipped learning \citep{grzesiak2024}, reusable building blocks \citep{albuquerque2023}, chatbots~\citep{nagel2024}, hands-on approaches \citep{konig2024}, and project-based learning (PrBL)~\citep{chow2021,revoredo2024, delgado2023}. We are particularly interested in PrBL because it has been widely used in undergraduate teaching and allows students to practice the application of theoretical content in projects (usually working in teams and interacting with real stakeholders) and, as a consequence, students can be more prepared for the job market.  

PrBL is an active learning approach based on practical experiences. According to \cite{krajcik2006}, this approach allows students to obtain knowledge and skills by working over an extended period investigating and solving a problem. The problem is often connected to the outside world, and students are encouraged to collaborate, communicate, and think critically to solve it, applying knowledge previously acquired \citep{pan2023}. It requires students to manage their time, resources, tasks, and roles effectively. This approach may be applied in specific courses or throughout an entire curriculum. In addition, it can be carried out by individuals or small groups, often supported by a team of teachers who act as advisers and consultants \citep{mills2003}.

\citet{padua2024} defined a PrBL-based educational framework that is grounded on the BPM life cycle and best practices for project management. Graduate students studying BPM in an administration course used the framework to analyze and prioritize improvements in company's business process. The results pointed to a significant improvement in the students' ability to apply theoretical concepts in a real environment, develop practical skills, and acquire a better comprehension of BPM, making them aware of the importance of end-to-end process management. Complementarily, \citet{revoredo2024} presented experiences and lessons learned about various teaching approaches applied in two editions of a BPM subject of a Master's course. In one edition, the authors used flipped classroom and PrBL, and in the other they adopted lecture-based instruction with in-class assessments and PrBL. In both editions, the goal was to improve a real-world project, which involved the redesign and validation of a to-be business process through interactions with stakeholders. Regarding the benefits of applying PrBL, the authors noted that the students~\citep{revoredo2024}: (i) learned about the business domain; ii) understood the relevance of BPM in real-world business; and (iii) increased their communication and organization soft skills.

In the same line of our study, \citet{delgado2023} presented a PrBL and hands-on approach to modeling business process integrated into a Computer Science curricula. Additionally, this approach also supports the development of process-driven systems. 

\subsection{Soft Skills in the Era of AI-assistance Software Development}
\label{sec:ai_era}
 
Aiming to deal with this new era of AI, organizations and educational institutions should adopt an integrated approach to training their future workforce, emphasizing digital and key competencies, such as creativity, communication, collaboration, and critical thinking \citep{pelaez-sanchez2024}.

In a complementary way, \cite{muzulon2025} observe that while technical expertise remains essential, soft skills—such as leadership, teamwork, communication, and adaptability—are increasingly critical for success in the digital era. In addition to that, a detailed understanding of the risks associated with using AI techniques, such as Large Language Models (LLMs), and related ethical aspects is essential for higher education graduates to maximize their employability \citep{wilkinson2024}.

In particular, in the context of software engineering, \cite{marlowe2026} highlight the need for Software Engineering Education Curricula that enlace process thinking, systems theory, and ethical engagement, integrated with generative AI technologies, and place greater emphasis on soft skills, including critical thinking, problem solving, creativity, reflective, intellectual flexibility, communication, teamwork, leadership, interdisciplinary or multidisciplinary thinking---the ability
to look at issues or problems from multiple perspectives---and the capacity for lifelong learning.

More precisely, in the AI age, a software engineer may serve less as a coding specialist and more as a collaborator, reviewer, supervisor, manager, or technologist, interacting with tools to develop, refine, and integrate parts of software systems \citep{marlowe2026}. In addition, these professionals must embrace the roles of a designer, shaping user-centric solutions, and an interpreter, bridging the gap between complex technical systems and business goals \citep{marlowe2026}. It is worth noting that the growing importance of hybrid profiles for software engineers lies in integrating technical expertise with socio-emotional capabilities \citep{nakash2025}. 
 
To the best of our knowledge, there are few studies that report experiences on PrBL specifically to teach business process modeling in computing-related undergraduate courses. There is also a scarcity of studies that investigate in-depth hard and soft skills exercised by students in such subject that could help foster essential competencies for future computing professionals. 

\section{Method}\label{sec:report}

This section presents the method adopted in this study. It is based on an industry-driven, project-based learning approach to teach business process modeling to computer science students (undergraduates in the second semester of a software engineering course) of a large public university during the second semester of 2024. The \textbf{main learning objective} of this study was to involve students in real-world situations, increase their learning experience, and make them face different job market challenges, aiming to practice relevant hard and soft skills for their career.

We prepared a feedback form\footnote{The complete form is available at: [Not available due to the double-blind review process.] 
It includes informed consent.} to collect data about the participants' profiles and opinions concerning hard and soft skills exercised throughout the semester. This form contained 5 questions for collecting student profile information and 5 and 11 questions for gathering the hard and soft skills, respectively (see Table \ref{tab:formQuestions}). To collect data regarding the skills we adopted a 5-point Likert scale (completely agree to completely disagree). The student could also provide any other comment regarding each skill. We conducted a pilot study with five students from another class to validate the form and ensure that the questions were clear and unambiguous. We then adjusted the form considering the suggestions for improvement.

\begin{figure}[!h]
   \centering
   \includegraphics[width=145mm]{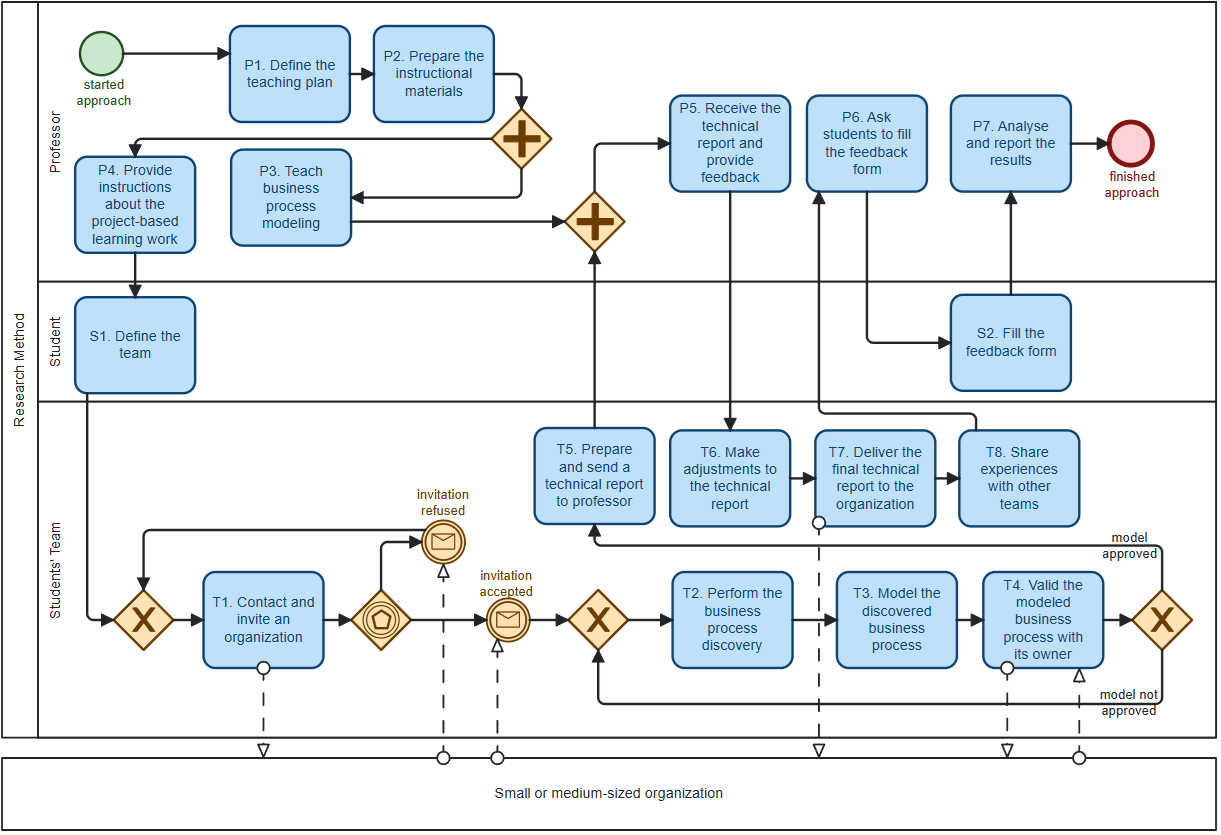}
   \caption{Stages of the method}
   \label{fig:research_method}
\end{figure}

Figure \ref{fig:research_method} presents the stages of the method that is composed of stages performed by professor (P1 to P7), student (S1 to S2), and student's team (T1 to T8). 

Firstly, the professor responsible for the business process modeling subject defined the teaching plan (\textbf{P1}), as shown in Table \ref{tab:teachingPlan}, according to the 
learning objective previously defined and previous experience with teaching theory and practice in this subject. This plan encompassed one-semester classes with 4  class hours per week (2 hours for theoretical content and 2 hours for practical exercises) and 4 hours outside class per week (when students performed their activities in the companies). Topic I (in this table) was worked as theoretical classes, while the others as theoretical and practical classes, including examples, case studies, and exercises, all of them supported by tools (e.g., bpmn.io\footnote{\url{https://bpmn.io/}}). Following this, the professor prepared the instructional materials according to the teaching plan (\textbf{P2}) and taught business process modeling  using corresponding materials (\textbf{P3}).  

\begin{table}[H]
    \small
    \centering
    \caption{Feedback questions}
    \begin{tabular}{|p{2cm}|p{10cm}|}
    \hline \textbf{Category}  & \textbf{Question}  \\\hline\hline
    I. Students' profile &	What is your course? \\\hline
    & What semester are you studying?  \\\hline
    & What is your previous experience (in years) in practical software systems development?	\\\hline
    & Which roles have you played in the development of software systems? \\\hline
    & What is your level of knowledge about the domain (in which your team modeled the business process)? \\\hline
    II. Hard skills  & a) The theoretical content given in the classroom was fundamental for me to understand how the organization's business process modeling should be conducted. \\\hline
    & b) The experience allowed me to live the role of a business analyst, interacting with the owner or participants of the business process. \\\hline
    &  c) The experience allowed me to understand each element of BPMN notation that could be used to represent each part of the business process model appropriately.  \\\hline
     & d) The experience taught me how to use a CASE tool (such as bmpn.io) to build business process models. \\\hline
    &  e) The experience allowed me to understand the importance of business process models for building software systems.
  \\\hline
    III. Soft skills  & f) Initiative: The experience helped me train my ability to take the initiative to seek to understand the organization's business increasingly. \\\hline
    & g) Teamwork: The experience helped me train my teamwork ability. \\\hline
    & h) Communication: The experience helped me train my communication ability with team members and the process owner.  \\\hline
      & i) Motivation: The experience allowed me to practice my motivation to deliver a business process model aligned with the process owner's expectations. \\\hline
    &  j) Proactivity: The experience helped me train proactivity, carrying out the tasks necessary to deliver the business process model within the expected deadline.   \\\hline
      & k) Professionalism: The experience allowed me to have a more professional attitude since I was interacting with representatives of a real organization. \\\hline
    &  l) Collaboration: The experience allowed me to train collaboration, contributing with other team members to understand the organization's business and represent it in a business process model. \\\hline
       & m) Decision Making: The experience encouraged me to train my ability to make decisions throughout  the business process modeling, for instance, to identify the elements of the notation, to schedule additional meetings to understand the business process better, etc. \\\hline
    &  n) Leadership: The experience enabled me to train my ability to lead my team to understand the organization's business and represent it in a business process model.  \\\hline
      & o) Agility: The experience allowed me to train my ability to be agile and perform the tasks necessary to build the organization's business process model. \\\hline
    &  p) Resilience: If there was an unforeseen situation, the experience enabled me (or would enable me) to train my ability to seek other ways to perform the tasks necessary to build the organization's business process model.  \\\hline
    \end{tabular}
    \label{tab:formQuestions}
\end{table}

In the first class, the professor also made available a detailed description of the project for the students\footnote{The complete description is available at: [Not available due to the double-blind review process.]} 
(\textbf{P4}). This description explained the required content in each section of the report (delivered by students at the middle of the semester): (i) \textit{company description}: its historic information, main company field, customer niche, company size in terms of physical structure and number of employees, reason(s) for choosing the company for the project, list of company's business processes; (ii) \textit{business process chosen}: description of the business process, reasons for choosing it, process owner and participants, roles of participants, knowledge level of students on the business domain, \textit{as-is} business process model (before the validation with its owner), improvements in the model pointed out by the owner, and \textit{to-be} business process model (after improvement); (iii) \textit{discovery methods}: description of the methods adopted to discover and comprehend the business process (such as interviews, observation, document analysis, and workshops~\citep{dumas2013}), as well as when and how they were applied and involved participants; and (iv) \textit{self-evaluation}: the positive and negative points, challenges, and difficulties observed by each member of the team (i.e, students). Teams of up to five students could perform this project, and they grouped themselves accordingly (\textbf{S1}).

\begin{table}[h]
\small
    \centering
    \caption{Teaching plan for business process modeling subject using an industry-driven, project-based approach}
    \begin{tabular}{|p{2cm}|p{10cm}|}
    \hline \textbf{Topic}  & \textbf{Description}  \\\hline\hline
    I. Introduction to BPM &	This topic presents the theoretical aspects of BPM (importance, benefits, efforts, and costs involved), and gives an overview of the five key activities (modeling, analysis, improvement, enactment, and monitoring) for the BPM \citep{weske2019}.  \\\hline
    II. Discovery techniques & This topic presents suitable methods and techniques to support the discovery of business processes widely adopted in the industry (e.g., interviews, observation, document analysis, event logs from information systems that automate business processes, and workshops).  \\\hline
    III. Essential elements for business process modeling & This topic presents and exemplifies the essential elements (e.g., activities - task and sub-process, events, data object, gateways, lane, pool, data flow, message flow, sequence flow) and basic diagrams (i.e., process diagram, collaboration diagram) from BPMN notation \citep{omg2014,white2008}. Besides, it presents the modeling best practices according to  \citet{mendling2010}.	\\\hline
    IV. Advanced elements for business process modeling  & This topic deals with the presentation and demonstration of process decomposition, process reuse, event handling
    exception handling, business rules, and advanced diagrams (choreography diagram and conversation diagram). \\\hline
    V. Case studies and practical exercises & This topic crosscuts most others and proposes case studies and practical exercises to simulate real situations that need to model the business to comprehend the organization better. \\\hline
    VI. Supporting tools & To engage the students and show them the importance of supporting tools to increase both productivity and models' quality and reduce syntax issues, it is important to present and always adopt modeling tools throughout the subject (e.g., bpmn.io, Bizagi\footnote{\url{https://www.bizagi.com}}).  \\\hline
    \end{tabular}
    \label{tab:teachingPlan}
\end{table}

Subsequently, each team contacted different organizations and selected one for their project (\textbf{T1}). For that, the organization needed to provide information about its business and would receive the resulting business process model at the end of the semester. Next, each team performed the discovery of the most relevant and strategic business processes of the organization (\textbf{T2}), modeled it (\textbf{T3}), and validated the process model until the organization approved it (\textbf{T4}). These stages were conducted over the semester and contributed to a real immersion of students into the organization's environment and understanding of the organization's business process. Following, each team prepared and delivered a report containing the details of the project (\textbf{T5}). The professor analyzed the reports and provided feedback regarding improvements in the model and adjustments in the reports (\textbf{P5}). Each team then made the required adjustments (\textbf{T6}) and delivered a final report to the participating organization (\textbf{T7}). Later, in an experience-sharing stage, each team shared its experiences with others, highlighting the positive and negative aspects and lessons learned (\textbf{T8}). Next, the professor asked students to individually fill out the feedback form (\textbf{P6}). Most students completed the feedback form (45 out of 53) and provided informed consent, i.e., around 85\% (\textbf{S2}). Finally, the professor analyzed the collected data to identify the main findings
(\textbf{P7}), as detailed in Section \ref{sec:findings}.  

\subsection{Participants' Profile}\label{profile}

\begin{figure*}[!h]
   \centering
    \includegraphics[width=135mm]{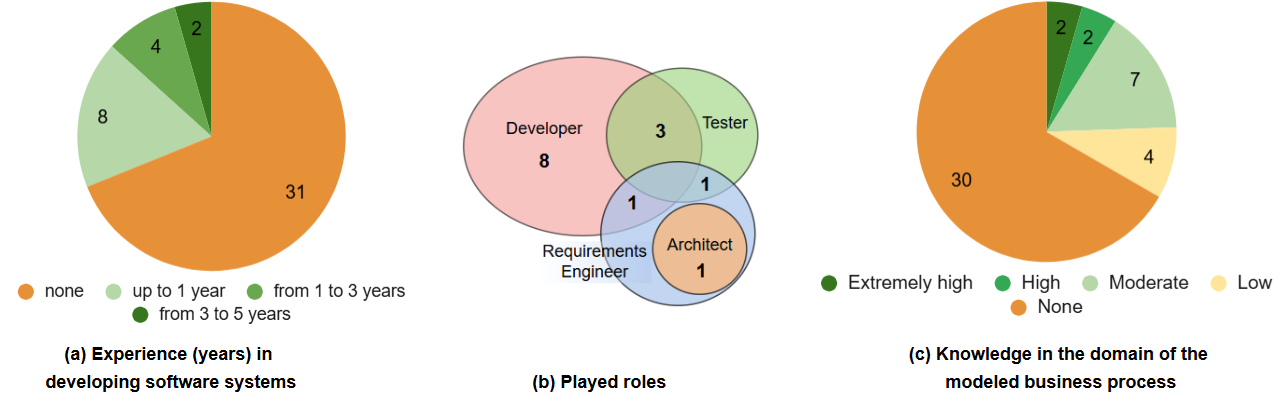}
   \caption{Students' profile}
   \label{fig:profile}
\end{figure*}

Figure \ref{fig:profile} summarizes the profile of the 45 students who answered the feedback form. Considering that the subject is offered in the second semester of their course, most students (i.e., 31) did not have experience in software development (Figure~\ref{fig:profile}.a). The eight students with up to one year of experience were already doing internships in companies. In addition, four others with one to five years of experience were students who completed technical training in software development  and were already working in the software industry. Considering the 14 students with experience in software development (Figure \ref{fig:profile}.b), most of them (i.e., 12) worked as developers, while a few performed other roles including requirements engineer, architect, or tester. As expected, as the students were in the beginning of their course, they performed roles that required less experience in software development. Regarding the knowledge of students in the domain of the business process that they managed, as Figure \ref{fig:profile}.c shows, most students (i.e., 34) had no or low previous knowledge. The other 11 students selected companies that: (i) they had contact  because they were customers consuming its products or services; or (ii) the companies' owners were relatives or friends of the students. We believe these reasons facilitated the communication between students and participating companies.

\section{Results}\label{sec:results}

This section presents the results\footnote{All collected raw data is available at: [Not available due to the double-blind review process]} 
regarding hard skills (Section~\ref{hard}) and soft skills (Section~\ref{soft}) trained by the students.

\subsection{Hard Skills}\label{hard}

\begin{figure*}[!h]
   \centering	\includegraphics[width=135mm]{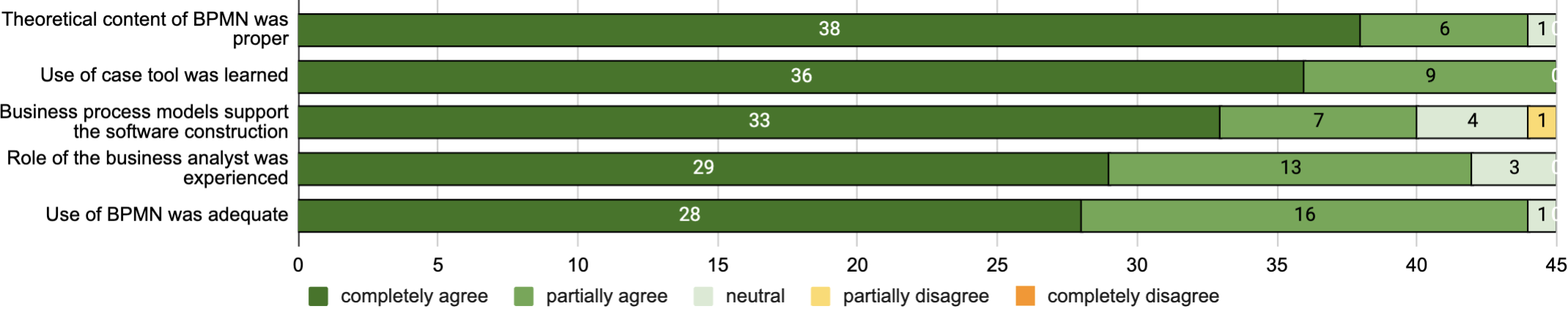}
   \caption{Feedback from students about their hard skills exercised}
   \label{fig:hardSkills}
\end{figure*}

Figure \ref{fig:hardSkills} summarizes the results regarding the hard skills exercised by students throughout the industry-driven project-based method. As expected, almost all students (i.e., 44 of 45) agreed (completely or partially) that the \textbf{theoretical content} given in the classroom was fundamental to understanding how to model the organization's business process. We believe this result is because, after each theoretical class, another class was used to solve practical exercises with a CASE tool. The following statements of students supported this view: S1 (\textit{``Doing exercises in each class helped a lot to model the process in the project.''}), S16 (\textit{``I really enjoyed doing exercises during class, it helps with fixing the content.''}), and S43 (\textit{``The exercises proposed weekly helped to consolidate the content.''}). 

All students agreed that the experience allowed them to learn the hard skill related to the \textbf{use of CASE tools} (such as bpmn.io) to build business process models, which aligns with previously discussed results. Some students' statements helped us to better understand this skill: S2 (\textit{``Learning to model [using a case tool] during classes and with exercises was vital [for my team] to  develop the project successfully.''}) and S43 (\textit{``The bpmn.io tool is great, easy to use, and efficient to support business process modeling''}).

Concerning the \textbf{importance (perceived by students) of business process models for software development}, most students (40 of 45) agreed that such models can support the development of software systems, as highlighted by several comments: S19 (\textit{``Undoubtedly, after this subject, I realized that modeling [business processes] is a crucial process for software development.''}) and S28 (\textit{``Especially after receiving the feedback from the professor [about our technical report], I understood the weight [of models] that was even greater than I had thought.''}). 4 students stayed neutral, and only 1 partially disagreed. We believe that were students still did not know the concepts related to software development because they were second-semester students, as confirmed by S23 (\textit{``I understand that to building software, it is necessary to follow a flow of activities, ordered step by step, as in a business process model. But apart from this comparison, I do not fully understand what software development is.''}). This statement to some extent aligns with the students' profile (Figure \ref{fig:profile}(b)) in which few students had experience in the software industry.  

As expected, most students (i.e., 42 of 45) agreed that they practiced the \textbf{role of a business analyst} as stated by S19 (\textit{``For me, it was incredibly interesting to understand how a company works up close. Experiencing things in practice was much more fun.''}), S28 (\textit{``It was very interesting, after collecting interview data and having this contact with a real company, setting up and deducing business processes became a rewarding experience.''}), and S43 (\textit{``It was an innovative experience, working with a real company, modeling and using [modeling and] discovery methods.''}). Nevertheless, 3 students stayed neutral. This could be justified because some students were responsible for writing the report instead of discovering and modeling business processes.

Almost all students (i.e., 44) agreed that the experience enabled them \textbf{to understand the elements of BPMN notation} that could be used to appropriately represent each part of the organization's business process model. Various statements from the students corroborated it, such as S10 (\textit{``All the BPMN elements [used] were taught to us and were practiced in class''.}) and S43 (\textit{``With the elements presented in class, there was no lack of any element in the modeling process that needed to be studied externally''.}). Only 1 student stayed neutral regarding this hard skill and affirmed that (\textit{``Modeling how activities should be updated (depending on customer feedback) was more difficult. I think we did not do an exercise that used the signal event with greater emphasis, what harmed the modeling regarding this semantic. But we made a mistake in not asking for help from the professor, as it would have been easier to solve.''}).

\subsection{Soft Skills}\label{soft}

\begin{figure*}[!h]
   \centering
   \includegraphics[width=135mm]{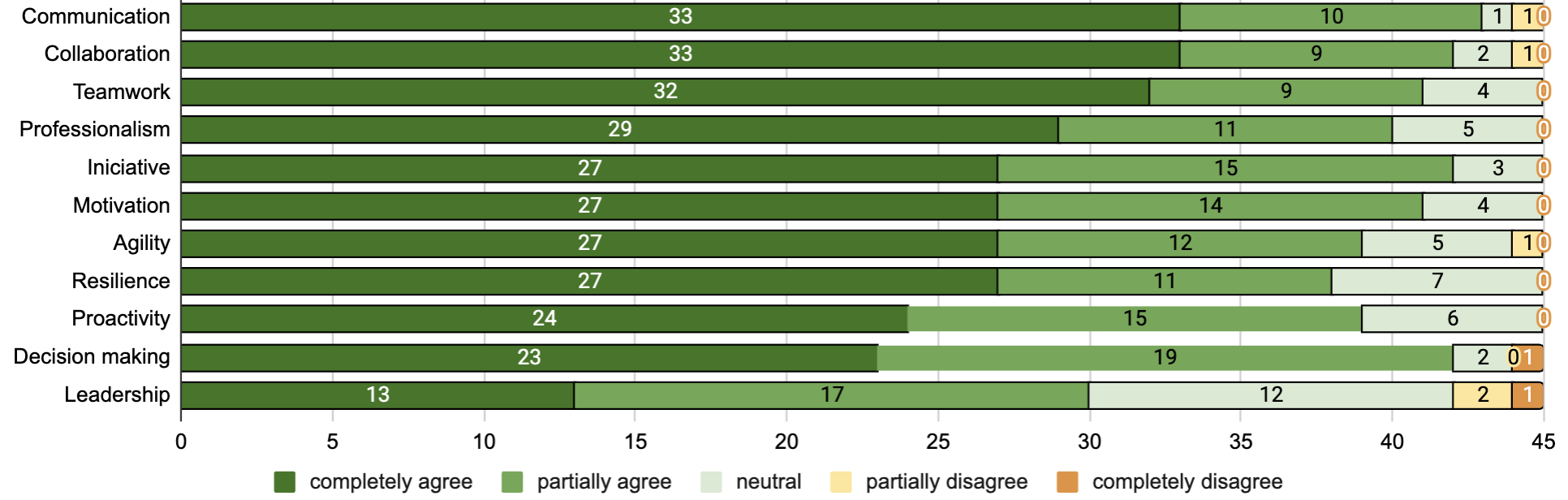}
   \caption{Feedback from students about their soft skills exercised}
   \label{fig:softSkills}
\end{figure*}

Figure \ref{fig:softSkills} summarizes the results related to the soft skills learned (or developed) by students.
\textbf{Communication}, \textbf{collaboration}, and \textbf{teamwork} were the soft skills most often reported, as also noted by student statements: S28 (\textit{``Talking, discussing, and deciding together made our work fruitful.''}) and S2 (\textit{``Teamwork is difficult because of the acquaintanceship and decisions that need to be made jointly. But it is vital to learn how to work together. In this sense, I enjoyed working in a team, and it provided development in this soft skill.''}). However, 2 students partially disagreed with the communication and collaboration as soft skills trained: S8 (\textit{``Unfortunately, we depend on good communication from others to work, and sometimes this is not the reality for everyone in the team.''}) and S2 (\textit{``I think collaboration was a barrier, since it took a long time to start the work and since we divided each part more, not collaborating much with each other.''}). Furthermore, as reported during the experience-sharing stage, some students mentioned the short time outside the classroom for the project, so they did not effectively collaborate to carry out the tasks.  

Surprisingly, 40 students agreed that they practiced \textbf{professionalism}, as stated by S6 (``\textit{In the meeting with the process owner, I stayed ahead of the questions, then I trained my formal communication skill.}'') and S39 (``\textit{It is a little of scary because only one people of my team is already working in an organization. This experience was really cool.}''). The 5 other students were neutral; we believe this is because they worked in a short time on the project as stated by S2 (\textit{``As I only had contact with the company's specialist once, I would say that the professional stance was neutral but effective.''}) and S8 (\textit{``Not completely yet, but in the future with more practice and seriousness on my part, yes.''}). Regarding the other two soft skills---\textbf{initiative} and \textbf{motivation}---they achieved the agreement of 42 and 41 students, respectively, as stayed by S2 (\textit{``My motivation was great, I think it was because modeling is an area that I like.''}) and S19 (\textit{``The professor is a very demanding person, and this helped me [motivated me] a lot in trying to do my best.''}). Furthermore, 3 and 4 students answered neutral for these skills, respectively. According to S28, \textit{``I feel that [initiative and motivation] are points that are most lacking in my person; I plan to improve this for the future.''}).

Thirty nine students agreed that they exercised \textbf{agility} (such as S2, \textit{``I think that leaving the work to be done much later than the defined start point greatly developed the soft skill of agility.''}). Similarly, 38 experienced \textbf{resilience}. Similarly, 39 students agreed that they exercised  \textbf{proactivity} because they were involved in executing the fundamental tasks 
to deliver value to process owners. One of the students (S6) stated that \textit{``I volunteered to go to the company and coordinate the meetings. I also volunteered to coordinate the team and separate the tasks.''}.

Remarkably, 42 students agreed they exercised the \textbf{decision making}. This is possibly due to several decisions made throughout the project. In particular, these decisions were essentially related to choosing the appropriate BPMN elements, selecting companies to be invited, finding an adequate schedule for meetings with the process owner, and distributing the project's tasks among the team members. However, 1 student completely disagreed (i.e., S2) and affirmed that \textit{``[he/she] was not prepared to make these decisions [due to their inexperience in using BPMN in practice].''}.

As expected,  \textbf{leadership} was the least exercised soft skill when compared to others, with only 30 students who agreed completely or partially. This can be justified because often only one member of each team took the leadership role. In addition, 12 students were neutral and 3 students partially or completely disagreed, as declared by S2 (\textit{``I did not have a leader during the execution of the work, all the activities were well divided [among the members].'')}. 

The students also reported other soft skills that they believed were exercised: \textbf{time management} (S8, S12, S14, S15), \textbf{patience} (S8, S17, S22),  \textbf{empathy} (S17, S32), \textbf{organization} (S14, S24), \textbf{adaptability} (S12), \textbf{self-confidence} (S37), and \textbf{logical reasoning} (S41). This experience shows us that students understand the importance of soft skills and develop them, as S15 stated: \textit{``I believe that the subject influenced us to overcome barriers in some areas, such as communication, professionalism, time management, and organization at the beginning of our undergraduate course.''}

In summary, this experience 
provided students with an opportunity to immerse in a company environment and practice various hard and soft skills.  These latter skills are aligned with those that are fundamental to the AI era, mentioned in Section \ref{sec:ai_era}. The next section discusses our main findings based on the results. 

\section{Discussion}
\label{sec:discussions}

This section discusses the main findings, the threats to validity of our work and countermeasure actions, and future directions in the field.

\subsection{Main Findings}
\label{sec:findings}

After conducting the research presented in this paper and  considering our previous experience in teaching business process modeling, our main findings are:

\begin{itemize}
     \item \textbf{The subject of business process modeling enables students to exercise soft and hard skills required for computing professionals in the age of AI}: As expected, soft skills necessary for computing professionals nowadays (Section \ref{sec:ai_era}), such as professionalism, decision-making, communication, collaboration, teamwork, initiative, leadership, and adaptability, 
     can be trained throughout the business process modeling subject. This subject can address the required competences of decision makers and can contribute to students further assuming fundamental roles in the AI era, such as manager, collaborator, designer, and interpreter \citep{marlowe2026}. If this subject is placed in the first year of their curriculum, it could also support the next steps in a student's academic life and  prepare them better for the job market, which is constantly evolving and increasingly requiring specific soft skills, such as those cited in Section \ref{sec:ai_era}. In addition, students can gain knowledge and practice in using BPMN and supporting tools properly, which are essential hard skills for managing business processes; hence, they can better understand the business as a whole and construct suitable and aligned software solutions to business goals, as mentioned in \cite{marlowe2026}. Furthermore, business process models are a source to elicit requirements of software systems that automate those businesses \citep{kraupvsa2024}; hence, the capacity to comprehend such models can be beneficial. 


    \item \textbf{Professors should adopt an industry-driven, project-based learning approach to teaching business process modeling:} Similar to other studies \citep{padua2024,revoredo2024,krajcik2006,pan2023}, we observed that adopting 
    an industry-driven, project-based learning approach seems to be more effective to achieve learning objectives. It allows students to immerse themselves in real business environments (gaining rich experience) and interact with professionals. We also observed the \textbf{organizations' openness to receive students and interest in gathering knowledge from academia}, i.e., such an approach can benefit both organizations and students. At the same time, students can then solve real-world problems in practice by applying concepts learned in theory. Hence, we observe \textbf{practical, real-world activities can motivate students}, who would be more motivated to learn when there are real-world projects and practitioners involved, as also stated in~\citep{padua2024,revoredo2024}. In the case of business process modeling, the students' motivation seems to have also  increased due to the organizations' openness and the  potential impact of the project results on the organizations; and

    \item \textbf{Relationships between academia and industry}: The knowledge and experience from academia, even that generated at the undergraduate level, can be transferred to the software industry. 
    This transfer is possible through individual researchers and informal collaboration, so it does not depend solely on government programs, large funded research projects, or the involvement of several researchers.
    We also observe that few companies are totally aligned with current technologies or have benefited from knowledge already spread in academia; hence, many companies can still benefit from academic knowledge, such as the modeling of business processes.

\end{itemize}

\subsection{Threats to Validity}
\label{sec:threats}

The main threats to validity of this work and countermeasure actions were:

\begin{itemize}
    \item \textbf{Feedback form:} A threat to validity might have been our feedback form construction, particularly regarding soundness questions. To mitigate this threat, we investigated the most common hard skills associated with business process modeling~\citep{dumas2013} and the most mentioned soft skills for competent professionals~\citep{borges2024}. We also conducted a pilot study involving five undergraduate computer science students to check the order and  clarity of the questions. Hence, we have greater confidence that our list of questions was complete and correct. 
    
    \item \textbf{Data collected:} Another threat to validity is related to the data collected because students could be concerned with their final grades, biasing their responses. To mitigate this threat, students were asked to respond to the questionnaires anonymously. Moreover, the professor's interest in their feedback was highlighted, regardless of whether it was positive or negative. Another threat might be the students' capability to remember details of their experiences, which might have negatively affected the thoroughness of their feedback. To mitigate this threat, the feedback form was filled out for each student after finishing the experience-sharing stage, minimizing the possibility  of students forgetting the details of the experience.

    \item \textbf{Interpretation of results:} The subjectivity of the interpretation of results, conducted by the researchers, might be another threat. To mitigate this threat, one of the authors carefully analyzed and interpreted the raw data (from the feedback form) based on years of experience in teaching business process modeling, while the other authors reviewed it. When there was disagreement among the authors, they discussed their opinions and reached a consensus. 
    
    \item \textbf{Generalization of results}: We cannot ensure the generalization of results because 
    this study involved specific details that could impact the results. Some examples of details are the student profiles, the characteristics of the organizations involved, and the workload of students during the semester. Despite this, the opinions of the students from the study may add value to future offerings of the subject.
    
    \item \textbf{Reproducibility and auditing:} The reproducibility and auditing of our study might be a threat to validity. To make our study reproducible to other researchers and mitigate this threat, we provide access to the project's description, the feedback form, and all raw data through an external material\footnote{URL: 
    [Not available due to the double-blind review process.]}.
    
\end{itemize}

\subsection{Future Directions}
\label{sec:future_work}

We now discuss some future directions that should be addressed in our own work, and in the computing education community:

\begin{itemize}

    \item \textbf{Conducting similar studies}: It is necessary to conduct similar studies to ours considering a larger number of students from different courses (such as computer science, computer engineering, and information systems) using other learning approaches (e.g., cases, gamification, and flipped learning) to compare to 
    industry-driven project-based learning approaches 
    and obtain statistically significant results. It is also necessary to collect feedback from participating organizations to gather evidence of the real benefits of experiences like this, particularly regarding knowledge and experience transfer.

    \item \textbf{Including business process modeling subject in computing-related courses:} Computing-related courses (such as computer science, computing engineering, information systems, and software engineering) should carry out an analysis of their course curricula to include business process modeling subject. This is due to the many benefits of this topic for students, particularly 
    the training of soft skills required in the AI age. 
    This subject should ideally be offered in the first semesters of the course or before (or concomitant with) subjects associated with software engineering, like requirements engineering, software design/modeling, and software architecture. This could help students recognize the importance of business processes and their modeling---which would enable a better comprehension of an organization and its business and the development of software systems to automate this business. This could also allow future professionals to act as interpreters when bridging the gap between technological solutions and the business level \citep{marlowe2026}.
    
    \item \textbf{Adjusting computing reference curricula}: Computing reference curricula should also encompass subjects related to BPM. These subjects not only make it possible for students to develop soft and hard skills but also prepare them to better understand organizational context, and comprehend, critically analyze, and even improve an organization's business processes before developing software systems that automate them. 

\end{itemize}

\section{Conclusion}
\label{sec:conclusion}

A significant distance between the business and technical levels still remains when developing software systems, although researchers and practitioners have increasingly mentioned BizDev \citep{moreira2023,fuentes2024}. Additionally, businesses have increasingly become more complex, in terms of growth, globalization, technological advancement, market condition, and so on. In parallel, developing software systems in the AI era requires hybrid profiles for computing professionals, considering technical and non-technical capacities, and an understanding of the risks and ethical aspects involved when using AI techniques. This paper calls attention to the importance of teaching the theory and practice of BPM, including process modeling and supporting tools, to leverage mainly soft skills of computing students, aiming to better prepare them for the job market. 

As with other undergraduate subjects, business process modeling can help students develop diverse soft skills essential in the era of AI-assistance software development, and we believe that those skills (such as professionalism and communication) that can be more effectively trained when in contact with real-world industrial settings can be more developed. Such skills and others practiced (such as teamwork, collaboration, proactivity, decision making, adaptability, and leadership) can also be relevant throughout a student's academic life, helping them to face the competitive labor landscape in the future. In addition, such a subject is  one that enables students to comprehend the meaning of ``business'' and its diverse concepts and aspects. Students with broader knowledge (not just in the theory and practice from computing subjects that the courses often provide) may be more valuable in industry. Hence, our takeaway message is: \textit{The sooner business process modeling is taught in undergraduate computing courses (for instance, using an industry-driven, project-based teaching approach), the sooner students can benefit during their course and professional careers}. We believe students would be better prepared to comprehend business processes more easily and even develop corresponding software systems that better fit business goals. 

\section*{Disclosure statement}
\begin{itemize}
    \item Conflict of interest: The authors declare no conflict of interests.

    \item Consent to participate: The feedback form, created to collect data about the participants' profiles and opinions regarding hard and soft skills exercised, contains a consent statement in the first page. The participants could agree or disagree to take part in the study and, hence, by continuing or not to fill out the questionnaire.

    \item Data availability: The supplementary artifact containing the feedback form, complete description of the project, and the raw data is openly available on GitHub at: \url{https://github.com/istela/ExternalMaterial_ExperienceReport}

    \item Author contribution: Maria Istela Cagnin: Writing-original draft. Elisa Yumi Nakagawa and Rick Kazman: Writing-review and editing.
\end{itemize}

\section*{Funding}
This work was funded by Federal University of Mato Grosso do Sul and Brazilian funding agencies CAPES (Finance Code 001), FAPESP (2023/00488-5, 2024/00329-7), and CNPq (313245/2021-5).

\bibliographystyle{apacite}
\bibliography{ref}

\end{document}